\documentclass[conference]{IEEEtran}
\IEEEoverridecommandlockouts
\usepackage{cite}
\usepackage{amsmath,amssymb,amsfonts}
\usepackage{algorithmic}
\usepackage{graphicx}
\usepackage{textcomp}
\usepackage{xcolor}
\def\BibTeX{{\rm B\kern-.05em{\sc i\kern-.025em b}\kern-.08em
    T\kern-.1667em\lower.7ex\hbox{E}\kern-.125emX}}

\usepackage{algorithm}
\usepackage{booktabs}
\usepackage{subcaption}
\usepackage{comment}

\begin{document}

\title{Circuit Partitioning for the Quantum Internet
\thanks{© 2025 IEEE.  Personal use of this material is permitted.  Permission from IEEE must be obtained for all other uses, in any current or future media, including reprinting/republishing this material for advertising or promotional purposes, creating new collective works, for resale or redistribution to servers or lists, or reuse of any copyrighted component of this work in other works.

Sponsored in part by the Bavarian Ministry of Economic Affairs, Regional Development and Energy as part of the 6GQT project.}
}

\author{\IEEEauthorblockN{Leo Sünkel}
\IEEEauthorblockA{\textit{Institute for Informatics} \\
\textit{LMU Munich}\\
Munich, Germany \\
leo.suenkel@ifi.lmu.de}
\and
\IEEEauthorblockN{Thomas Gabor}
\IEEEauthorblockA{\textit{Institute for Informatics} \\
\textit{LMU Munich}\\
Munich, Germany}
\and
\IEEEauthorblockN{Claudia Linnhoff-Popien}
\IEEEauthorblockA{\textit{Institute for Informatics} \\
\textit{LMU Munich}\\
Munich, Germany}
}

\maketitle

\begin{abstract}
In a quantum internet, quantum processing units (QPUs) with varying architectures and capabilities may be connected through quantum communication channels, enabling new applications such as distributed quantum computing (DQC), a paradigm in which multiple QPUs execute a single circuit. However, remote operations between QPUs are expensive as they require the creation and distribution of entanglement throughout the network. It is therefore crucial to assign qubits to QPUs and partition circuits in such a way that the overall communication between QPUs is minimized. In this paper, we apply and evaluate simulated annealing and an evolutionary algorithm for this problem. We consider quantum networks with 25 nodes arranged in different topologies and QPUs with varying qubit capacities. The circuits evaluated contain 50 and 100 qubits. We show that the different metaheuristics all significantly outperform the baselines by drastically reducing the communication cost by over 40\%.
\end{abstract}


\section{Introduction}
Quantum communication networks and particularly a quantum internet \cite{kimble2008quantum,caleffi2018quantum,singh2021quantum} in which QPUs of various architectures are connected allow to envision new forms of applications for the quantum era, including quantum key distribution (QKD) \cite{bennett2014quantum}, blind quantum computing \cite{broadbent2009universal}, and distributed quantum computing (DQC) \cite{caleffi2018quantum}. The main idea of DQC is to connect QPUs through classical and quantum communication channels to allow the collective execution of large circuits that single machines would not be able to achieve on their own \cite{caleffi2024distributed}. Thus, it can be viewed as a strategy to scale quantum computing to new dimensions. Various approaches or implementations have been suggested, including connecting QPUs in close proximity to form clusters or in large communication networks. Even further, the quantum internet is envisioned as a global network of QPUs with different architectures and capabilities. In this paper, we will focus on challenges that arise in DQC in networks with different QPUs arranged in various topologies. 

This paper is structured as follows. In Sec. \ref{sec:background} we briefly discuss the necessary background and present related work in Sec. \ref{sec:related_work}. We recap the problem of circuit partitioning in Sec. \ref{sec:circuit_partitioning} before discussing the experiments performed as part of this work as all presenting the results in Sec. \ref{sec:experiments}. We examine the findings in Sec. \ref{sec:discussion} and conclude the paper in Sec. \ref{sec:conclusion}.

\section{Background}\label{sec:background}
In this section, we will first recapitulate the basic building blocks necessary to build quantum networks and a quantum internet, before proceeding with the fundamentals of DQC.

\subsection{Quantum Networks and the Quantum Internet}
In quantum networks, QPUs are connected by both classical and quantum communication channels. However, as qubits are fragile and entanglement is a crucial ingredient necessary for important protocols, for example, quantum state teleportation, new challenges emerge to enable communication over long distances. In a quantum network, Bell pairs (maximally entangled qubits) must be created and distributed throughout the network. Quantum repeaters \cite{briegel1998quantum,van2013designing} can be placed throughout the network that perform entanglement purification and swapping. With entanglement purification protocols, weakly entangled qubits can be used to create highly entangled qubits, which are necessary for many protocols, though note that $k$ weakly entangled qubits are used to create $n$ highly entangled qubits where $n<k$. The entanglement swapping protocol enables to entangle two qubits without them interacting with each other through the use of a third party. For example, consider three parties, Alice, Bob, and Charlie. Alice and Bob each possess a qubit of a shared Bell pair, as do Bob and Charlie. Now Bob can perform the entanglement swapping protocol and project the entanglement onto Alice and Charlie, thereby entangling both without requiring them to interact with each other directly.
For the purpose of this work, we use the term quantum internet to refer to a network of QPUs where each has its own qubit capacity that is, each node represents a different QPU architecture. These characteristics will become relevant when assigning subcircuits to QPUs, and we will discuss this in more detail next.

\subsection{Distributed Quantum Computing}
In DQC, before a circuit can be executed by nodes in the network, qubits must be assigned to the QPUs and circuits partitioned. That is, the circuit must be divided into subcircuits which can then be executed by the respective QPU. However, in most cases, non-local gates are unavoidable and remote operations must be performed. A remote operation (i.e., non-local gate) is a multi-qubit gate where the control and target qubits reside in different QPUs; however, in this work, the CX gate is the only multi-qubit gate present in all circuits.  If a non-local gate must be performed, the qubits can be teleported such that they are located on the same QPU and the gate can be executed locally. Alternatively, the remote-CX protocol can be executed where no qubit must change its physical location. Both protocols require a Bell pair. A major advantage of the DQC paradigm is that multiple small QPUs can jointly run a circuit that none could run on their own.

\section{Related Work}\label{sec:related_work}
The interest in the task of circuit partitioning for quantum networks or, more specifically, DQC, has been growing in recent years, with various approaches being proposed and evaluated by the research community. For example, graph partitioning methods \cite{daei2020optimized}, hypergraph partitioning \cite{andres2019automated,cambiucci2023hypergraphic} and evolutionary algorithms \cite{sunkel2024applying} have been applied to the problem. An evolutionary algorithm and simulated annealing to optimize the qubit assignment, as well as an evolutionary algorithm to optimize the quantum circuit itself are evaluated in \cite{sunkel2025time}. In \cite{nikahd2021automated} the authors apply a window-based approach, and in \cite{baker2020time} a time-slice approach is proposed. In \cite{mao2023qubit} the authors propose a heuristic local search as well as a hybrid simulated annealing approach. The approaches proposed so far focus on minimizing the overall communication cost. While approaches that take the underlying network topology into account when assigning qubits to QPUs have been discussed and evaluated in the literature, the actual QPU architecture is often irrelevant in the optimization. We address this problem in this work in that we consider quantum networks consisting of QPUs with different qubit capacity. We will formulate and discuss the problem in the next section.

\section{Circuit Partitioning}\label{sec:circuit_partitioning}
The objective of the approaches in this work is to assign qubits to nodes (i.e. QPUs) in a quantum network (or a quantum internet) that results in minimal communication cost measured in number of required teleportations and remote-CX gates. In the evaluated setting, QPUs have different capacities such that a naive qubit assignment may result in unnecessary teleportations. Qubits are assigned a QPU for each time step of a circuit, resulting in a two-dimensional schedule, i.e., qubit assignment as proposed in \cite{sunkel2024applying} and illustrated in Table \ref{tab:example_assignment}. This assignment can then be used for circuit partitioning to create all subcircuits for each QPU. Usually circuits are partitioned and qubits assigned to QPUs such that the number of remote operations is minimized, as discussed in Sec. \ref{sec:related_work}. However, other constraints may also need to be considered in real networks when QPUs from different vendors or architectures, and thus capabilities, are connected to collectively run a quantum circuit. Moreover, in a potential quantum internet, not all QPUs will necessarily be directly connected to every QPU in the network, that is, QPUs may be arranged in arbitrary topologies, which also should be considered in the cost function.
The objective used in this work, which is based on \cite{sunkel2024applying,sunkel2025time}, is formally defined as:

\begin{align}\label{eq:cost_fn}
    \min_{x} \sum_{t=0}^{T-1} \sum_{(i,j)\in CX_t}^{}  \textnormal{dist}(x_{i,t}, x_{j,t}) \\  \notag
    + \sum_{t=0}^{T-2} \sum_{q=0}^{Q-1} \textnormal{dist}(x_{q,t}, x_{q,t+1}) + \lambda
\end{align}

\noindent where $CX_t$ denotes all CX gates at time step $t$, $x_{i,t}$ the QPU assigned to qubit $i$ at time step $t$, and $\textnormal{dist}$ the distance in number of hops between QPUs; $q$ also denotes a qubit. Additionally, a penalty $\lambda$ is added in each time step if a QPU is assigned more qubits than allowed by its capacity. $\lambda$ was set to 10000 in the experiments to avoid invalid solutions. 
Thus, the objective is to assign each qubit to a particular QPU in the network for each time step of the quantum circuit such that the overall communication cost is minimized. Each QPU has a different capacity, i.e., a limited number of qubits reserved for computation. A wide range of optimization and metaheuristic algorithms can be applied to these problems. In this paper, we apply simulated annealing (SA) and an evolutionary algorithm (EA) and compare the results to two static qubit assignment strategies; we introduce these approaches next.

\begin{table}[b]
    \caption{Example qubit assignment to QPUs. At each time step, every qubit is assigned to a QPU represented by an integer. If qubits change their assignment from one time step to the next, a teleportation must be performed. Similarly, if a multi-qubit gate is performed at a time step where the involved qubits are located at different QPUs, a remote-CX gate must be performed. Both approaches consume a Bell pair and thus can be weighted equally in a cost function.}
    \label{tab:example_assignment}
    \centering
    \begin{tabular}{|l|c|c|c|c|}
    \hline
        & \multicolumn{4}{c|}{\textbf{Time}} \\ \hline
        \textbf{Qubits} & $t_0$ & $t_1$ & $t_2$ & $t_3$ \\ \hline
         $q_0$ & 0 & 0 & 1 & 1 \\ \hline
         $q_1$ & 1 & 1 & 0 & 0 \\ \hline
         $q_2$ & 0 & 1 & 1 & 0 \\ \hline
         $q_3$ & 1 & 0 & 0 & 1 \\ \hline
    \end{tabular}
\end{table}

\subsection{Static Qubit Assignment}
With a static qubit assignment, we refer to a strategy in which qubits are fixed to a QPU throughout the entire circuit execution. Remote operations are solely performed by the remote-CX protocol and no qubits are moved through teleportation. We evaluate two static assignments in this work that will also function as baselines. In the first method, which we will refer to as ''Successive Assignment'', each QPU is filled up with successive qubits until its capacity is reached. For example, QPU 0 with a capacity of three would be assigned the first three qubits of the circuit. QPU 2 with a capacity of five with the next five qubits, and so forth until all qubits are assigned to a QPU. In the second method, qubits are assigned to QPUs in descending order of qubit capacity, that is, starting with the QPU with the highest capacity, qubits are successively assigned to QPUs. Again, each QPU is filled up before the next QPU is selected according to its capacity. We refer to this method as the ''Capacity Based Assignment''. Note that the total capacity of the network, that is, the sum of all QPU capacities is greater than the qubits contained in the circuits. Therefore, not all QPUs must be fully occupied.

\subsection{Simulated Annealing}
SA is a metaheuristic algorithm that iteratively operates on a single solution. Its main components are a current solution, temperature $t$, a cooling rate $\alpha$, a cost function, and a function that creates a neighbor, i.e., similar solution. In each iteration, the currently accepted solution is slightly modified to create a neighbor. Whether the neighbor is accepted as the new solution depends on its fitness (cost) and current temperature. That is, the probability of accepting a new solution is formally defined as \cite{kirkpatrick1983optimization}:

\begin{equation}
    p(n, c, t) = \begin{cases}
        1 & \text{if } n < c, \\
        \textnormal{exp(-(n-c)/t)} & \text{else}
    \end{cases}
\end{equation}

\noindent where $n$ is the fitness or cost of the neighbor and $c$ of the current solution, and $t$ the current temperature. Thus, better solutions are always accepted; however, the algorithm allows for exploration by accepting worse solutions with a probability determined by the current temperature. More specifically, with high temperature the chances of accepting worse solutions is higher. SA usually starts with a high temperature and gradually decreases according to the cooling schedule. Therefore, in the early phase of the algorithm, when the temperature is high, SA can explore more solutions that are worse than the current solution. However, as the algorithm progresses and the temperature drops, the probability of accepting worse solutions decreases, leading to more exploitation than exploration. A high-level overview of SA is shown in Alg. \ref{alg:sa}. 

\begin{algorithm}[tb]
    \caption{Simulated Annealing}
    \label{alg:sa}
    \begin{algorithmic}
        \REQUIRE $initial\_temp$
        \REQUIRE $cooling\_rate$
        \REQUIRE $initial\_solution$
        \REQUIRE $max\_iterations$
        \STATE $c \leftarrow initial\_solution$
        \STATE $best \leftarrow c$
        \STATE $t \leftarrow initial\_temp$
        \FOR{$i \leftarrow 0$ to $max\_iterations$}
            \STATE $t \leftarrow$ $t$ * $cooling\_rate$
            \STATE $n \leftarrow create\_neighbor(c)$
            \IF{$\text{cost}(n) < \text{cost}(c)$ \textbf{or} $\exp\left(-\frac{(\text{cost}(n) - \text{cost}(c))}{t}\right) > \text{random}(0,1)$}
                \STATE $c \leftarrow n$
            \ENDIF
            \IF{\text{cost(c)} $<$ \text{cost(best)}}
                \STATE $best \leftarrow c$
            \ENDIF
        \ENDFOR
        \RETURN $best$
    \end{algorithmic}
\end{algorithm}

\subsection{Evolutionary Algorithm}
EAs are metaheuristic optimization algorithms based on principles of biological evolution \cite{holland1992adaptation,eiben2002evolutionary}. An EA is a population-based and iterative approach in which a number of individual solutions compete for selection to participate in the next generation, i.e., iteration of the algorithm. New individuals, i.e., solutions, are created by combining elements of ''parent'' individuals through crossover operations. Individuals can further be modified by mutations, that is, functions that introduce slight random variations in solutions. A population is sorted by fitness, a scalar that determines how good a solution is. At the end of each generation, $n$ solutions are selected to participate in the next generation, while the rest are replaced by newly created solutions. A high-level overview of a simple EA is given in Alg. \ref{alg:ea}. The crossover and mutation methods are usually problem-specific. Different replacement strategies exist and in this work we use an elitist approach in which the worst $k$ individuals are replaced by $k$ ''children'', i.e., the newly created individuals (offspring).
The EA used in this work is inspired by the approach proposed in \cite{sunkel2024applying}, and the crossover and mutation methods were implemented accordingly. 

\begin{algorithm}[tb]
    \caption{Simple Evolutionary Algorithm}
    \label{alg:ea}
    \begin{algorithmic}
        \REQUIRE $initial\_population$
        \REQUIRE $mutation\_rate$
        \REQUIRE $replace\_rate$
        \REQUIRE $o\_rate$ \COMMENT{Offspring rate}
        \REQUIRE $generations$
        \STATE $population \leftarrow initial\_population$
        \FOR{$i \leftarrow 0$ to $generations$}
            \STATE $parents \leftarrow sample(population)$
            \STATE $offspring \leftarrow crossover(parents, o\_rate)$
            \FOR{$o$ in $offspring$}
                \IF{$random(0, 1) < mutation\_rate$}
                    \STATE $mutate(o)$
                \ENDIF
            \ENDFOR
            \STATE $sort(population)$
            \STATE $replacement(population, offspring)$
        \ENDFOR
    \RETURN $population$
    \end{algorithmic}
\end{algorithm}

\begin{figure*}[tb]
    \centering
    \begin{subfigure}{0.3\textwidth}
        \centering
        \includegraphics[scale=0.2]{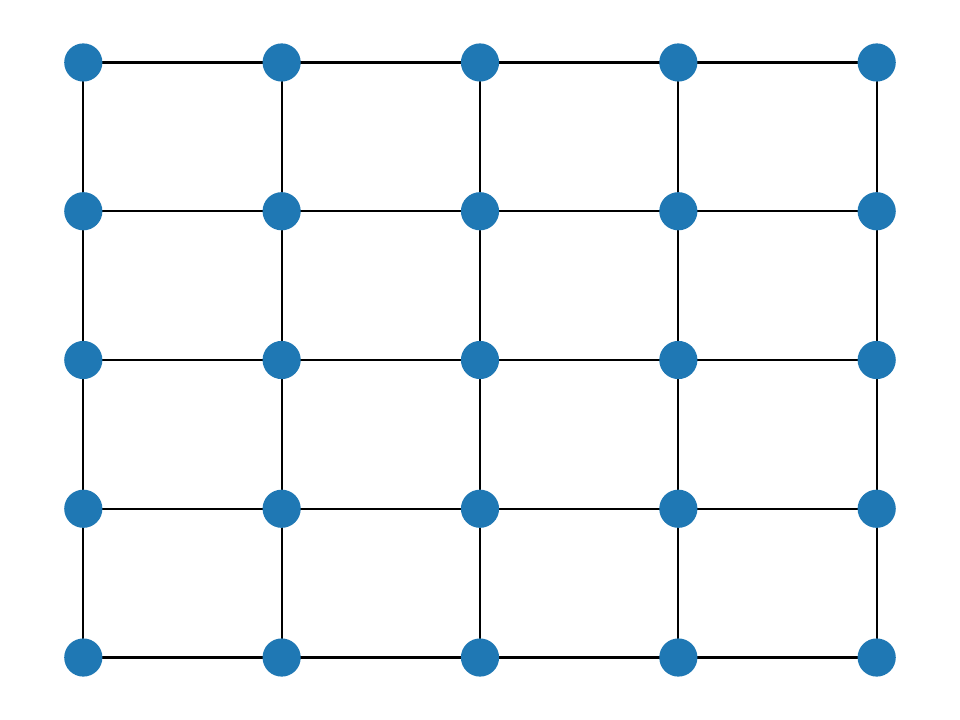}
        \caption{Grid network}
    \end{subfigure}
    \hfill
    \begin{subfigure}{0.3\textwidth}
        \centering
        \includegraphics[scale=0.2]{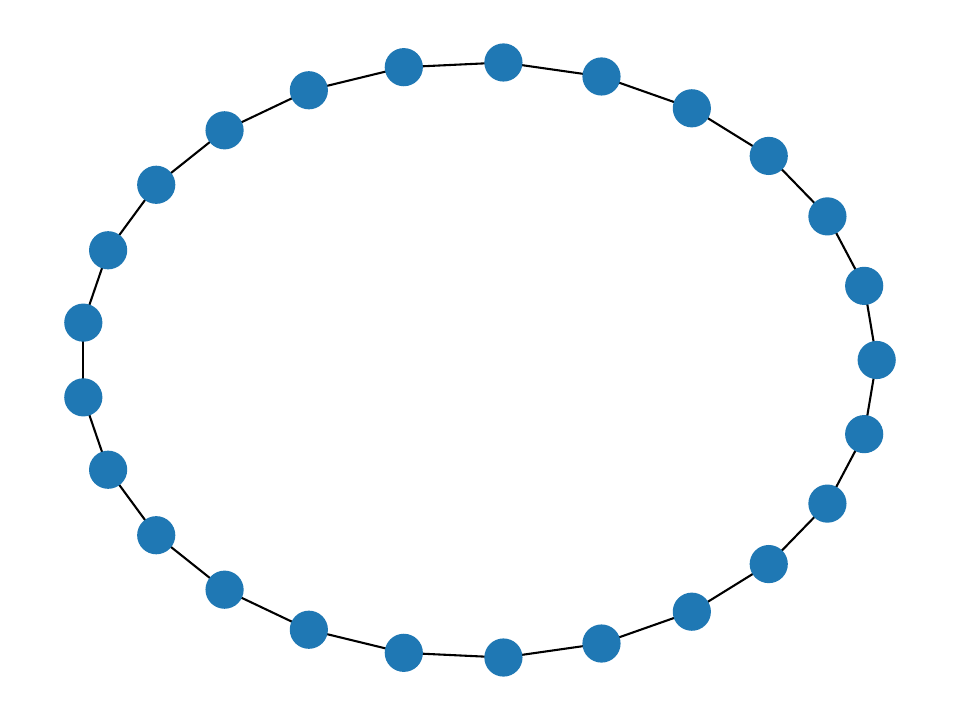}
        \caption{Ring network}
    \end{subfigure}
    \hfill
    \begin{subfigure}{0.3\textwidth}
        \centering
        \includegraphics[scale=0.2]{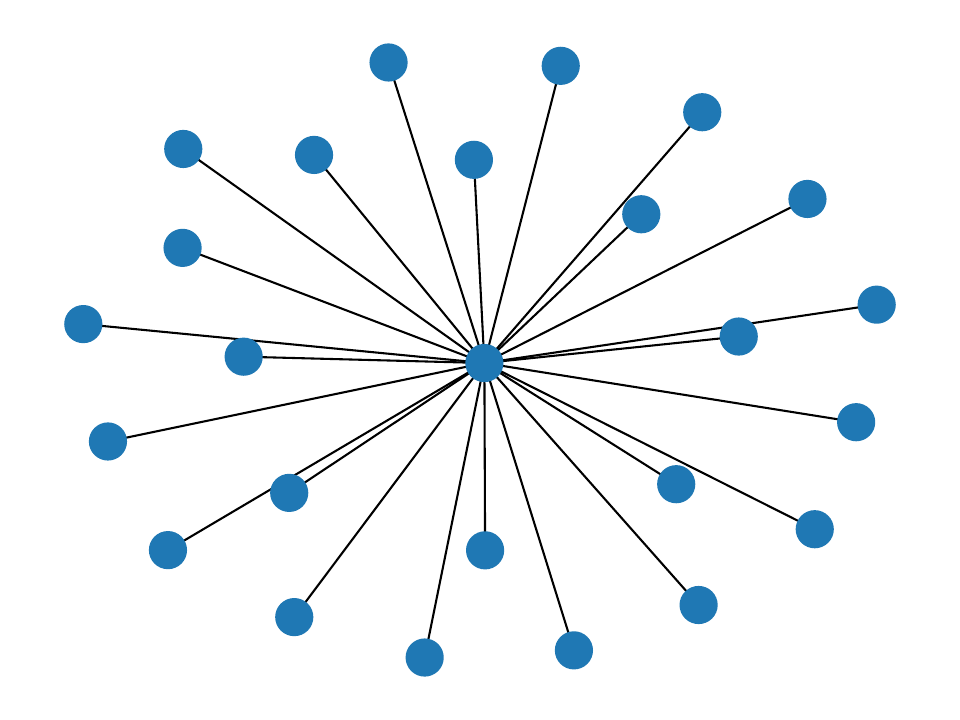}
        \caption{Star network}
    \end{subfigure}
    \caption{The topologies evaluated as part of this work. Each network has 25 nodes and the distance is measured in hops. }
    \label{fig:network_topologies}
\end{figure*}

\subsection{Approach}
We evaluate two different metaheuristic algorithms in this work. Each approach uses the same solution representation (as described above and illustrated in Table \ref{tab:example_assignment}) and fitness (i.e. cost) function defined in Eq. \ref{eq:cost_fn}. Furthermore, all proposed algorithms, i.e., the EA and SA use the same methods to create neighbor solutions; the only exception being the EA that additionally uses crossover to combine two solutions in order to create a new one. Moreover, while SA creates neighboring, i.e., similar solutions, the EA uses mutations to randomly alter offspring created through crossover; however, mutation and neighbor creation methods are identical. The approaches randomly choose one of the following mutation methods for each mutation/neighbor creation. Mutations can flip one or multiple cells, swap rows, columns, or two cells, and shuffle rows and columns in the solution matrix. The EA randomly chooses from two different crossover methods to create each new solution: (1) row-wise and (2) column-wise crossover. In both approaches, a random cut-off point is determined, and the new solution inherits all row/columns left/above this point from one parent and right/below from the other. The mutation and crossover methods used in this work are based on the approaches proposed in \cite{sunkel2024applying}. Using these methods, each algorithm proceeds as described above, and we compare them to the static assignment baseline methods.

\section{Experiments}\label{sec:experiments}
We start this section by giving an overview of the experimental setup and discuss how the evaluation of the proposed methods was performed. We then present the results of all experiments conducted as part of this work.

\subsection{Experimental Setup}
We evaluate the qubit assignment problem on quantum networks with different topologies (ring, star, and grid) consisting of 25 nodes (QPUs) with varying qubit capacities. The capacities were randomly determined and are listed in Table \ref{tab:network_config} and the networks used were created using NetworkX \cite{SciPyProceedings_11} and are shown in Fig. \ref{fig:network_topologies}. Circuits were randomly created using Qiskit \cite{qiskit2024} with 50 and 100 qubits. For the circuits with 50 qubits, each QPU's capacity was determined randomly with a value between 2 and 5, and between 2 and 10 for the 100 qubit circuit. The total capacity of the networks, that is, the combined capacity of all QPUs is more than required for the circuit. The total capacity of all QPUs is 89 for the circuit with 50 qubits and 146 for the 100 qubit circuit. Hyperparameters were determined experimentally and set for the EA as follows: the population size was 200, mutation rate 0.3, offspring (i.e., new individuals) for each generation was set to 60, and the algorithm ran for 5000 generations for the 50 qubit circuit and 10000 for the 100 qubit circuit. The starting temperature for SA was set to 10 and the cooling rate to 0.99. SA ran for 60000 and 100000 iterations for the 50 and 100 qubit circuits, respectively.

\begin{table}[tb]
    \centering
    \caption{The network and circuit configuration used in the experiments. Each QPU has its own qubit capacity.}
    \begin{tabular}{lcr}
    \toprule
        Nodes & 25  \\ \midrule
        Total Capacity Network 1 (50 qubit circuit) & 89 \\ \midrule
        Total Capacity Network 2 (100 qubit circuit) & 146 \\ \midrule
        Circuit50 CX Gates & 353 \\ \midrule
        Circuit50 Depth & 95 \\ \midrule
        Circuit100 Depth & 97 \\ \midrule
        Circuit100 CX Gates & 690 \\
    \bottomrule
    \end{tabular}
    \label{tab:network_config}
\end{table}

\begin{figure*}[tb]
    \centering
    \begin{subfigure}{0.3\textwidth}
        \centering
        \includegraphics[scale=0.3]{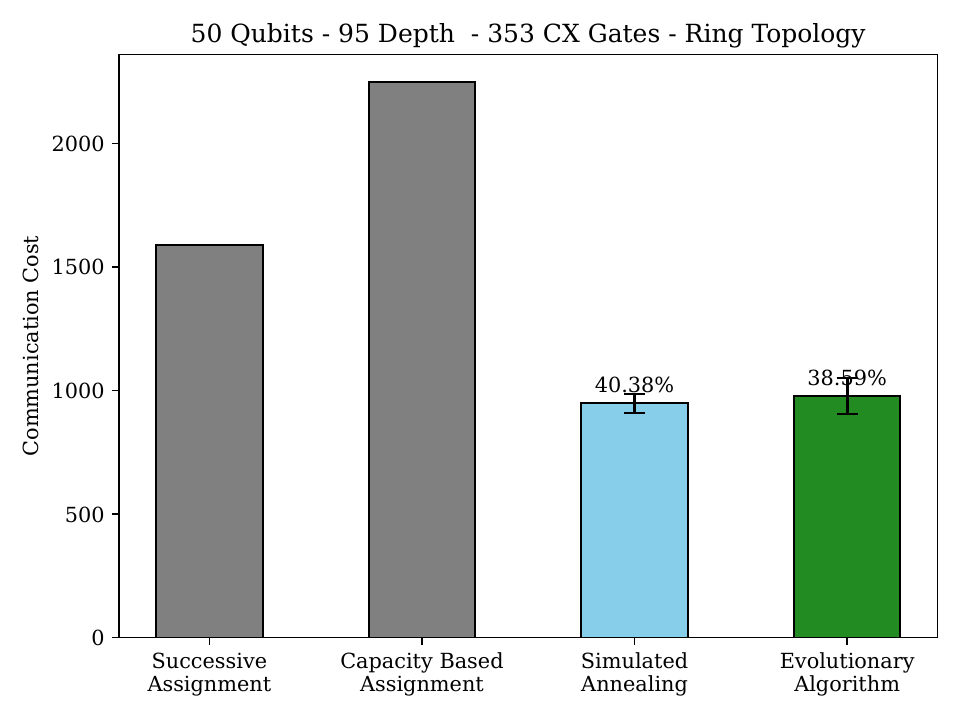}
        \caption{Ring Topology}
        \label{fig:50_qubits_ring_cost_comparison}
    \end{subfigure}
    \hfill
    \begin{subfigure}{0.3\textwidth}
        \centering
        \includegraphics[scale=0.3]{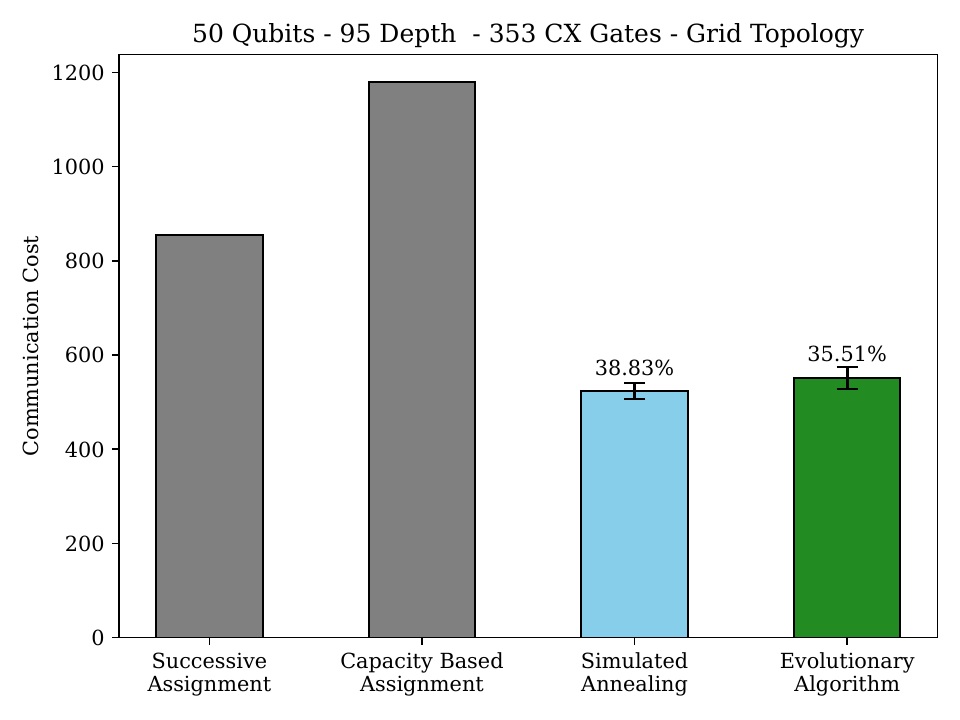}
        \caption{Grid Topology}
        \label{fig:50_qubits_grid_cost_comparison}
    \end{subfigure}
    \hfill
    \begin{subfigure}{0.3\textwidth}
        \centering
        \includegraphics[scale=0.3]{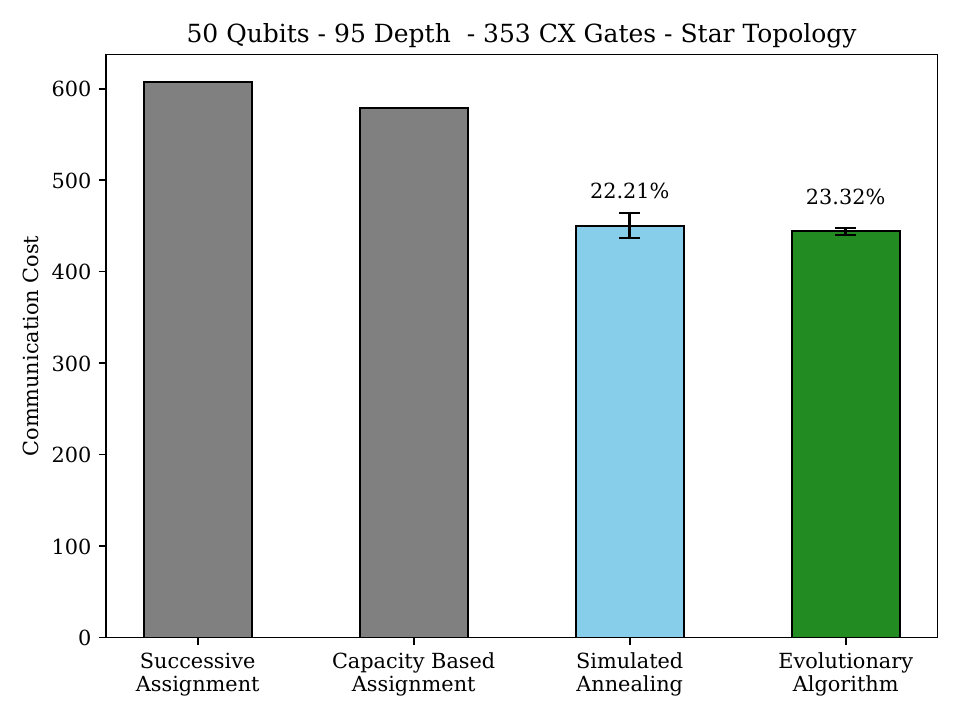}
        \caption{Star Top}
        \label{fig:50_qubits_star_cost_comparison}
    \end{subfigure}

    \begin{subfigure}{0.3\textwidth}
        \centering
        \includegraphics[scale=0.3]{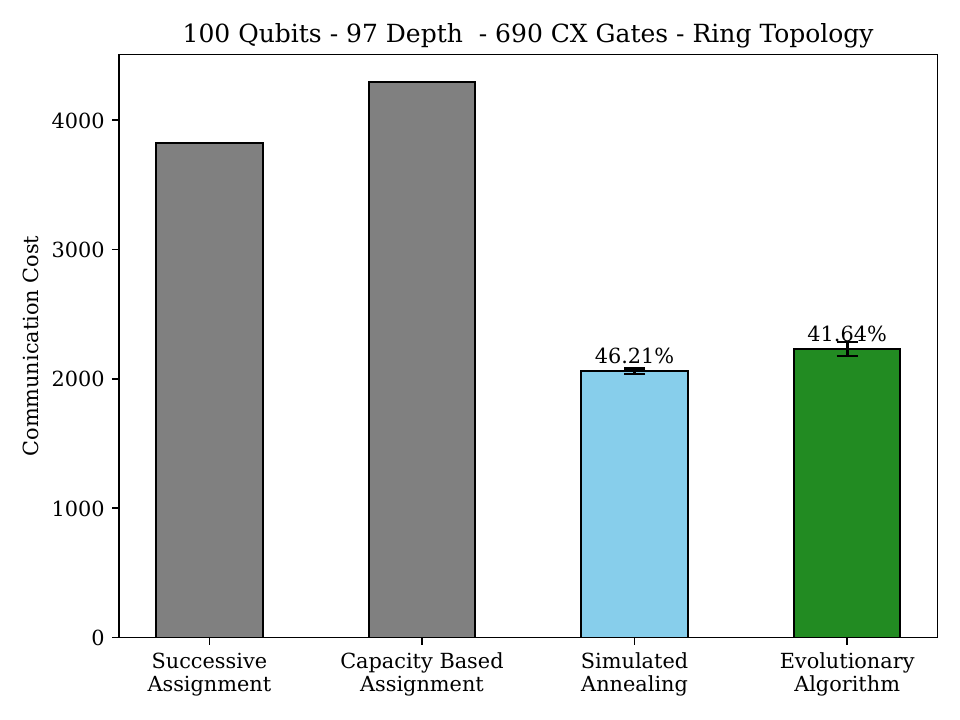}
        \caption{Ring Topology}
        \label{fig:100_qubits_ring_cost_comparison}
    \end{subfigure}
    \hfill
    \begin{subfigure}{0.3\textwidth}
        \centering
        \includegraphics[scale=0.3]{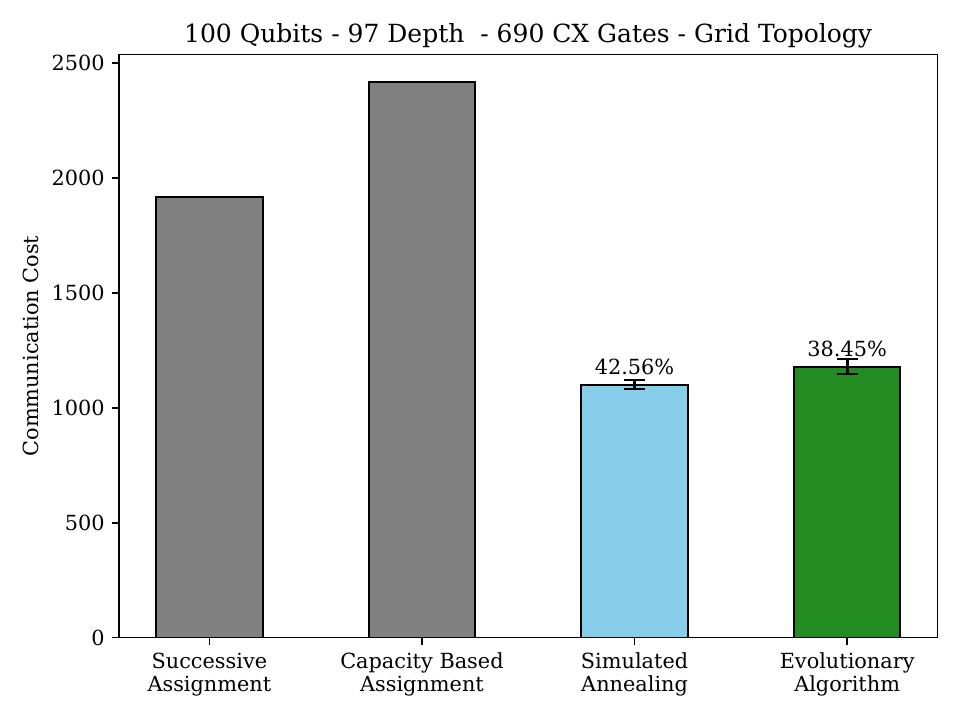}
        \caption{Grid Topology}
        \label{fig:100_qubits_grid_cost_comparison}
    \end{subfigure}
    \hfill
    \begin{subfigure}{0.3\textwidth}
        \centering
        \includegraphics[scale=0.3]{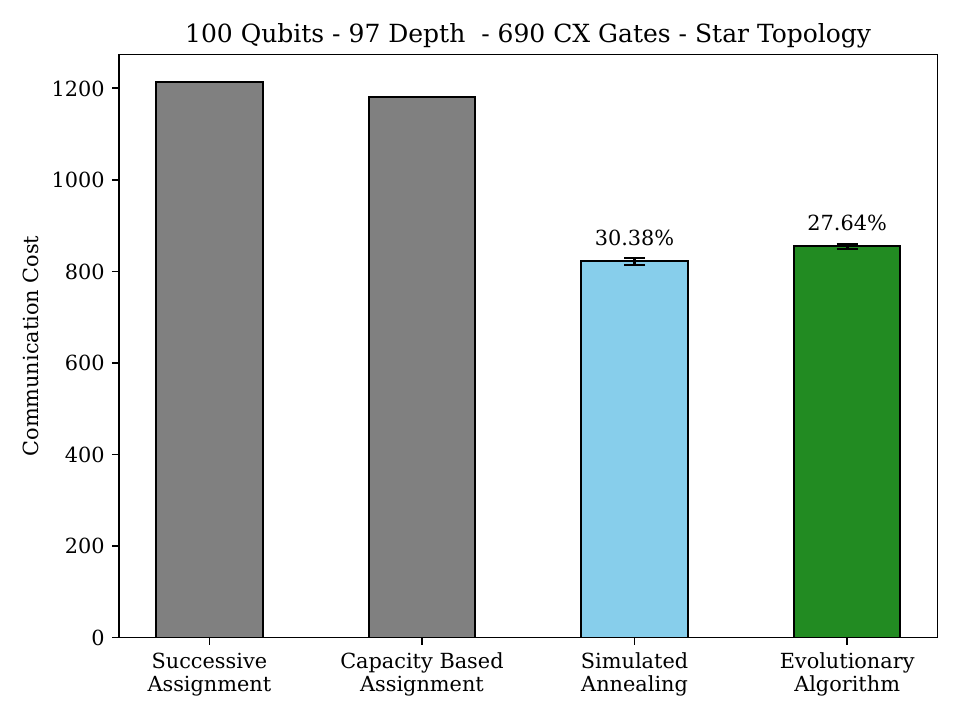}
        \caption{Star Topology}
        \label{fig:100_qubits_star_cost_comparison}
    \end{subfigure}
    \caption{Overview of the communication costs achieved for all experiments.}
    \label{fig:cost_comparison}
\end{figure*}

\begin{figure*}[tb]
    \centering
    \begin{subfigure}{0.45\textwidth}
        \centering
        \includegraphics[scale=0.45]{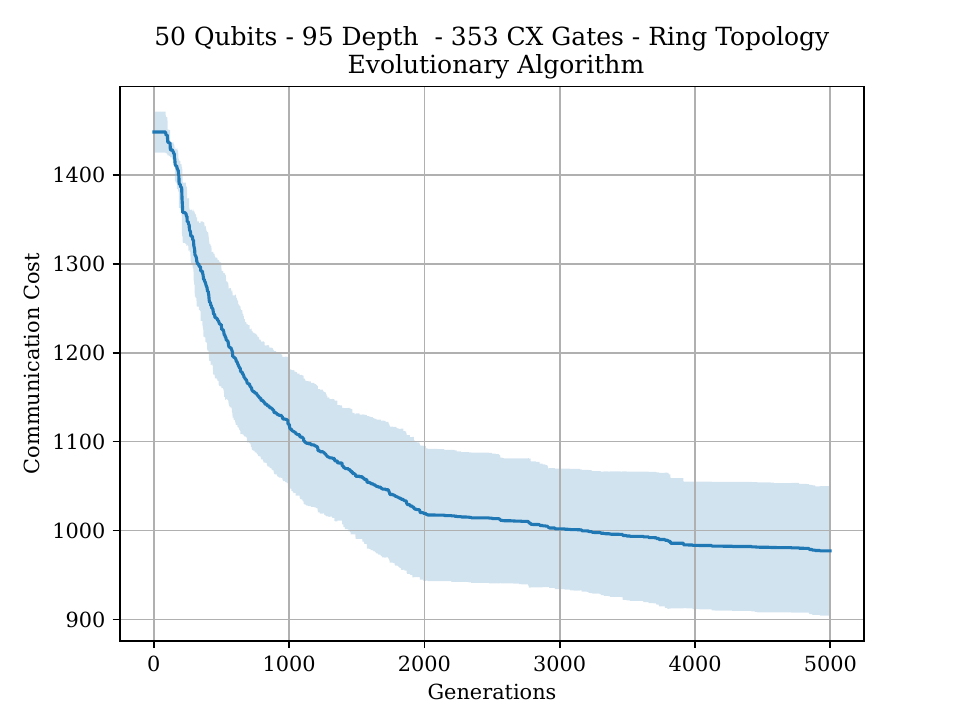}
        \caption{Evolutionary Algorithm - Ring Topology}
    \end{subfigure}
    \hfill
    \begin{subfigure}{0.45\textwidth}
        \centering
        \includegraphics[scale=0.45]{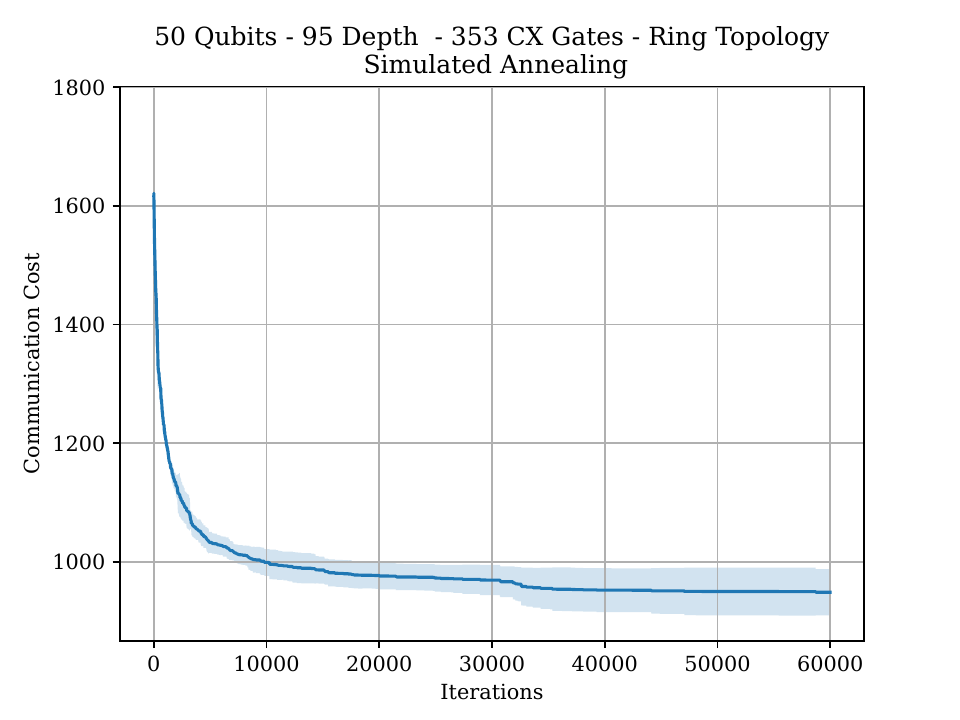}
        \caption{Simulated Annealing - Ring Topology}
    \end{subfigure}
    
    \begin{subfigure}{0.45\textwidth}
        \centering
        \includegraphics[scale=0.45]{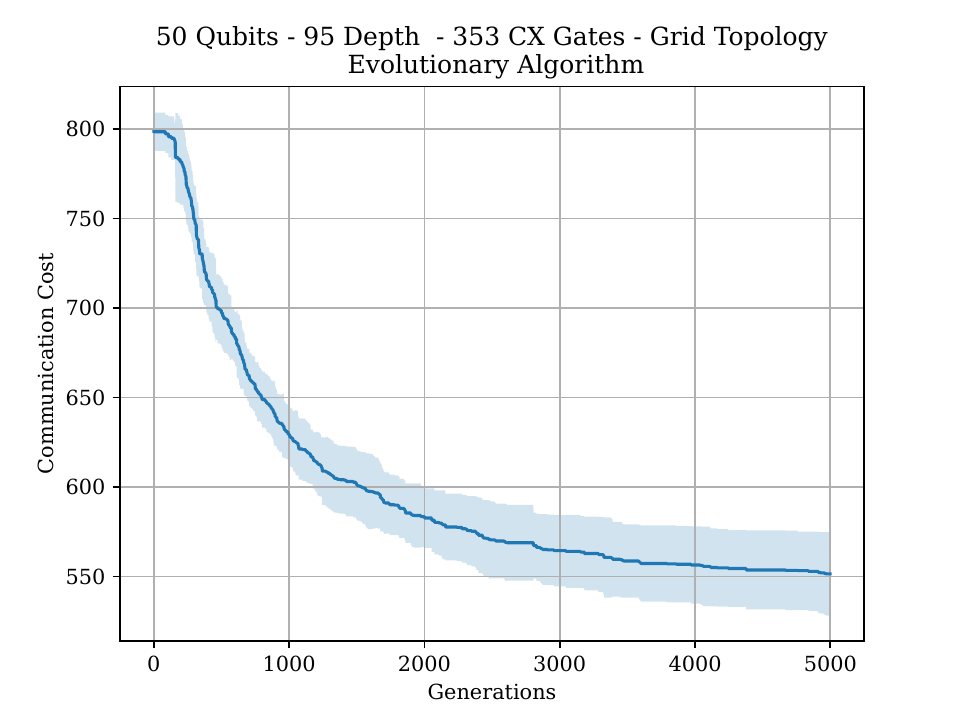}
        \caption{Evolutionary Algorithm -  Grid Topology}
    \end{subfigure}
    \hfill
    \begin{subfigure}{0.45\textwidth}
        \centering
        \includegraphics[scale=0.45]{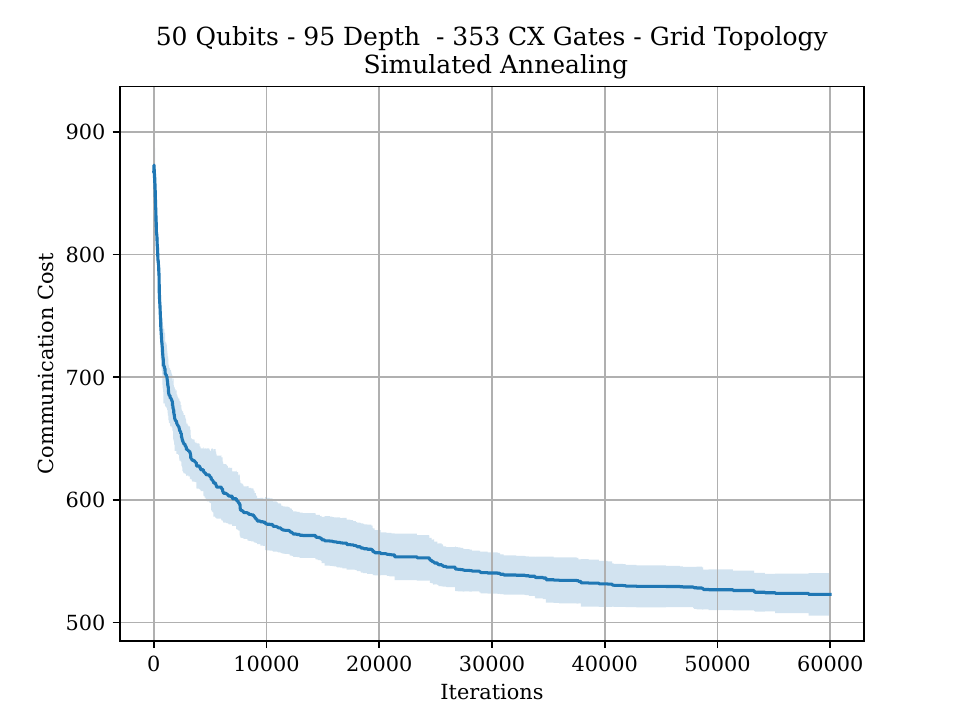}
        \caption{Simulated Annealing - Grid topology}
    \end{subfigure}
    
    \begin{subfigure}{0.45\textwidth}
        \centering
        \includegraphics[scale=0.45]{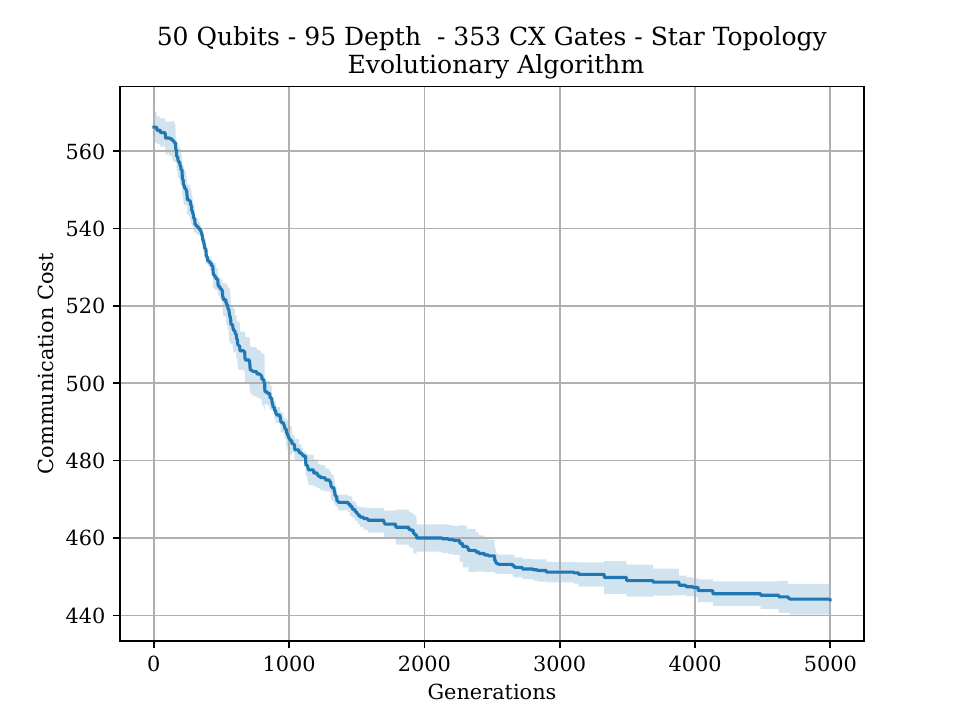}
        \caption{Evolutionary Algorithm - Star Topology}
    \end{subfigure}
    \hfill
    \begin{subfigure}{0.45\textwidth}
        \centering
        \includegraphics[scale=0.45]{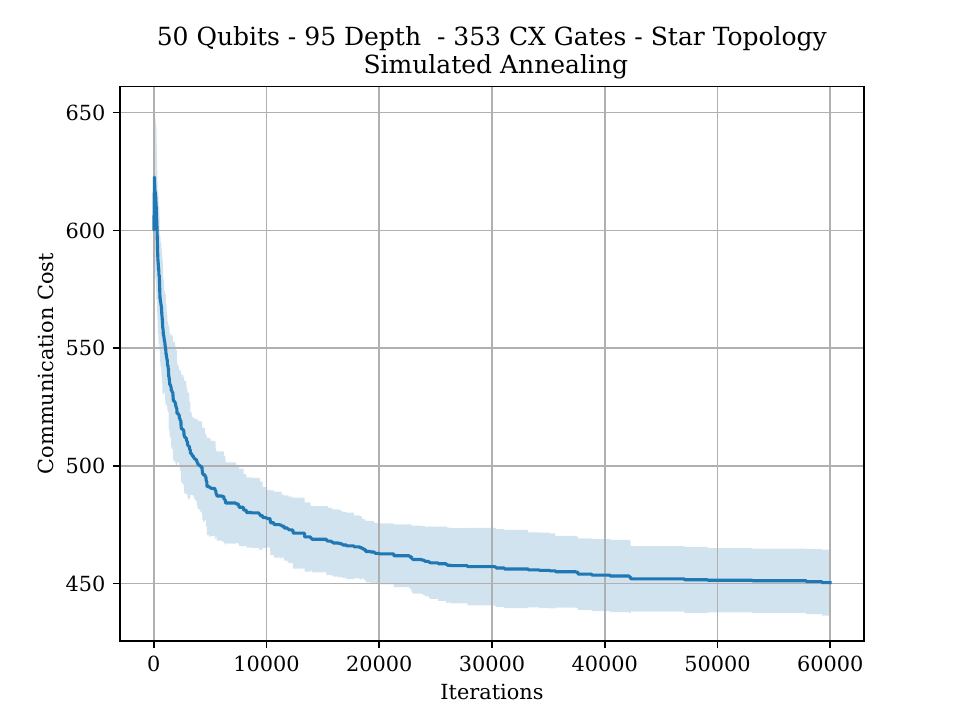}
        \caption{Simulated Annealing - Star Topology}
    \end{subfigure}
    \caption{Communication cost over the generations/iterations for the experiments on the circuits with 50 qubits. The left side shows the cost of the best individual of the EA in each generation; the right side the current cost in each iteration of SA. Plots show the mean over all seeds with standard deviation.}
    \label{fig:50_qubits_cost_over_iterations}
\end{figure*}

\begin{figure*}[tb]
    \centering
    \begin{subfigure}{0.45\textwidth}
        \centering
        \includegraphics[scale=0.45]{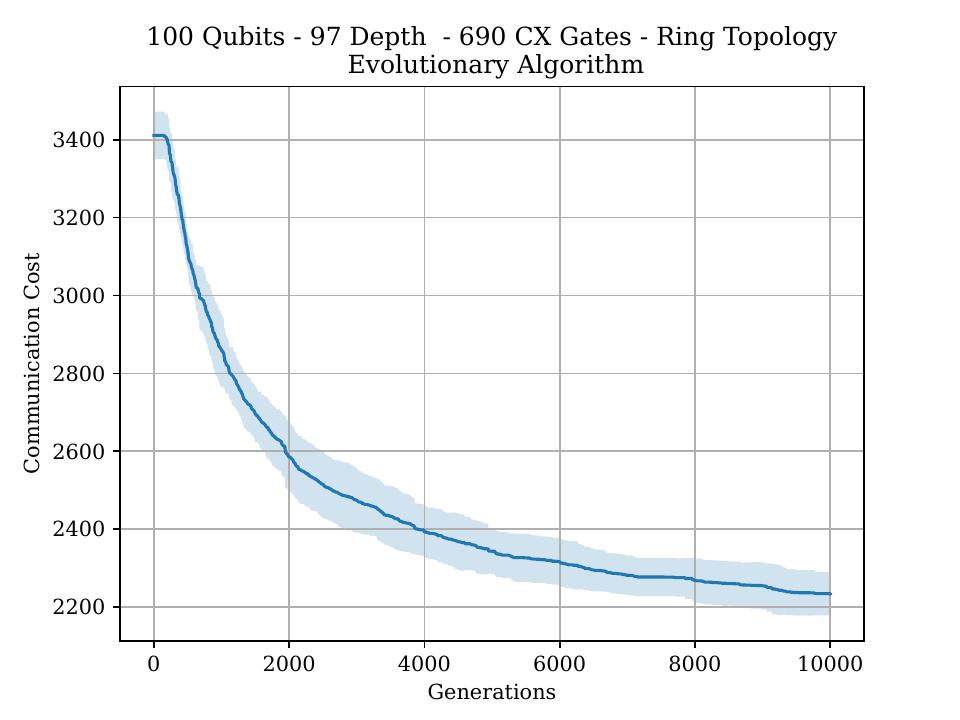}
        \caption{Evolutionary Algorithm - Ring topology}
    \end{subfigure}
    \hfill
    \begin{subfigure}{0.45\textwidth}
        \centering
        \includegraphics[scale=0.45]{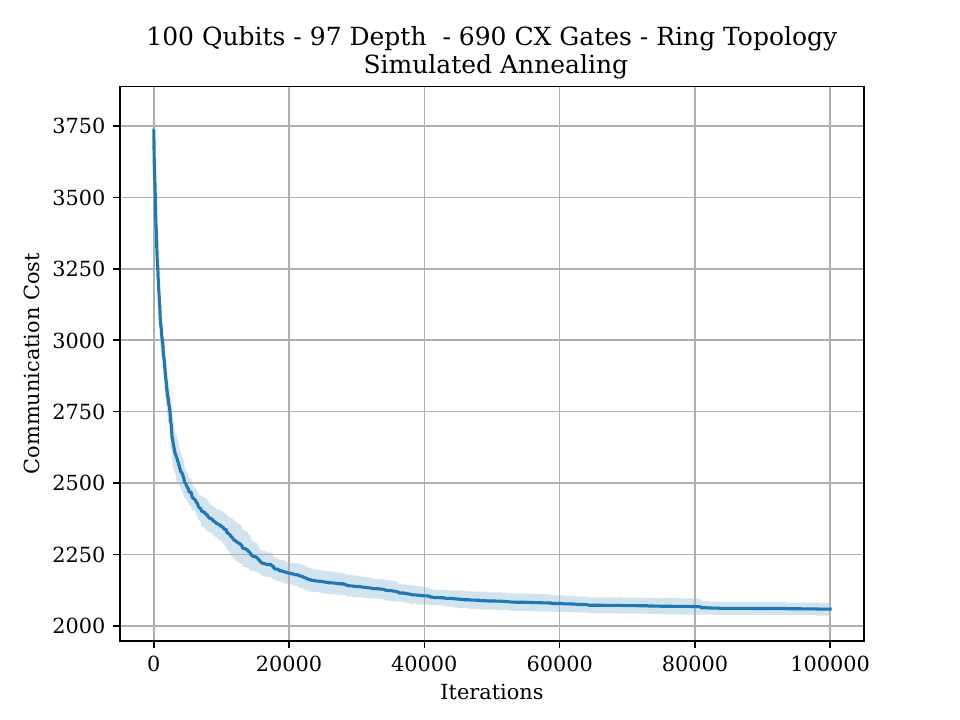}
        \caption{Simulated Annealing - Ring Topology}
    \end{subfigure}
    
    \begin{subfigure}{0.45\textwidth}
        \centering
        \includegraphics[scale=0.45]{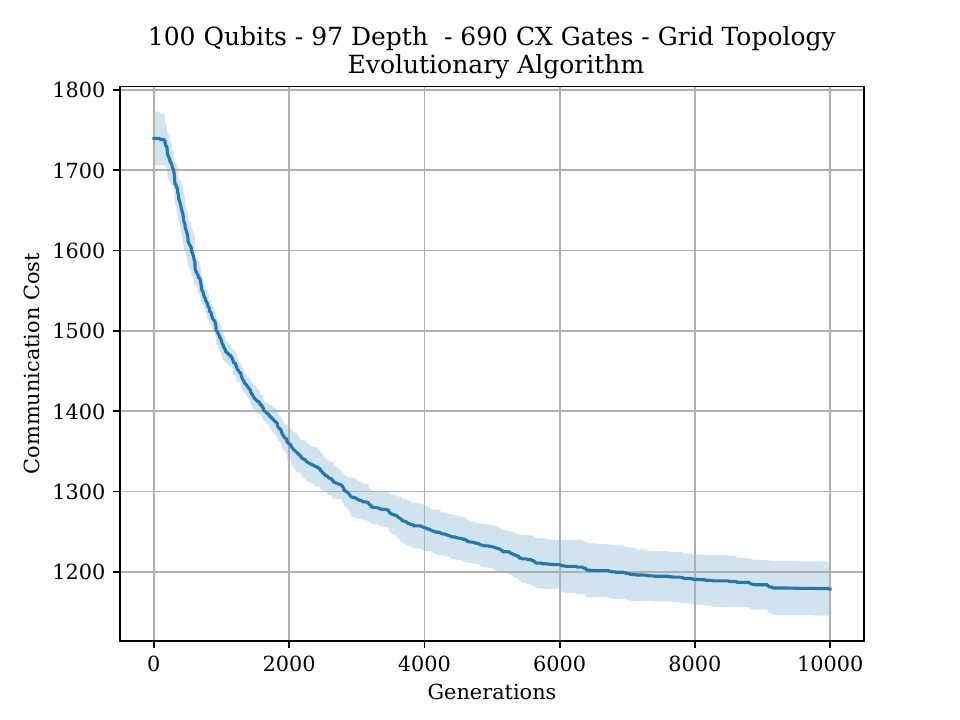}
        \caption{Evolutionary Algorithm - Grid Topology}
    \end{subfigure}
    \hfill
    \begin{subfigure}{0.45\textwidth}
        \centering
        \includegraphics[scale=0.45]{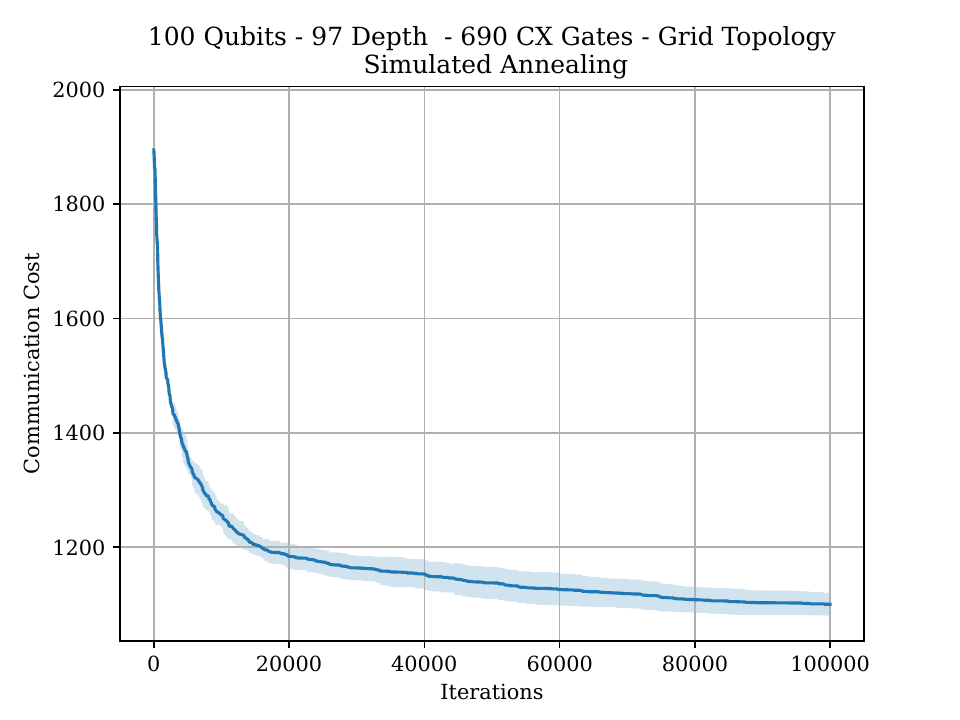}
        \caption{Simulated Annealing - Grid topology}
    \end{subfigure}
    
    \begin{subfigure}{0.45\textwidth}
        \centering
        \includegraphics[scale=0.45]{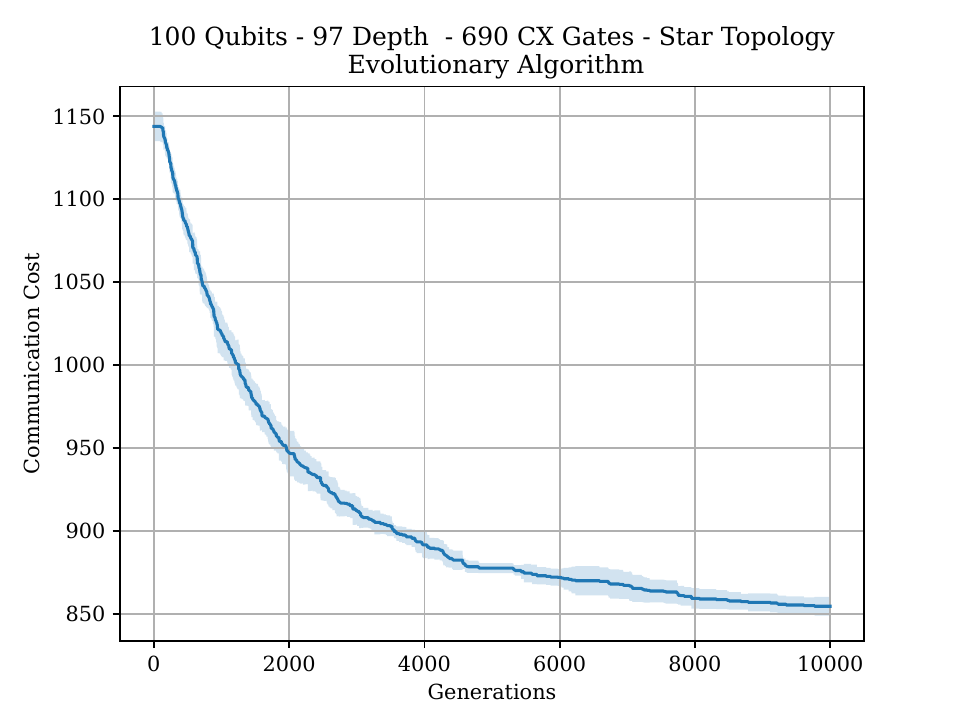}
        \caption{Evolutionary Algorithm - Star Topology}
    \end{subfigure}
    \hfill
    \begin{subfigure}{0.45\textwidth}
        \centering
        \includegraphics[scale=0.45]{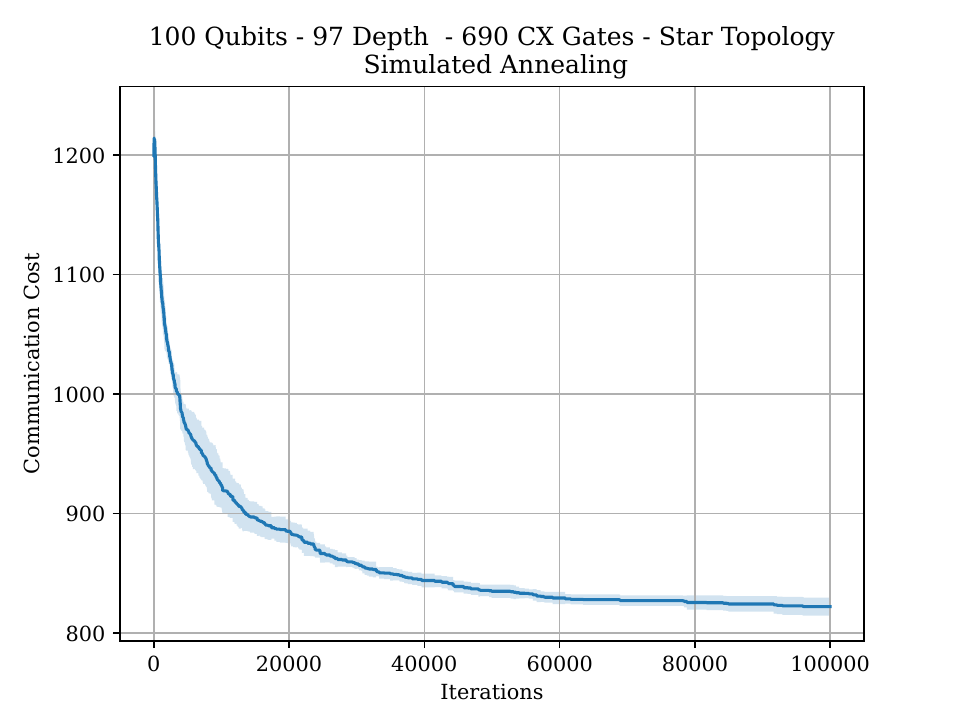}
        \caption{Simulated Annealing - Star topology}
    \end{subfigure}
    \caption{Communication cost over the generations/iterations for the experiments on the circuits with 100 qubits. The left side shows the cost of the best individual of the EA in each generation; the right side the current cost in each iteration of SA. Plots show the mean over all seeds with standard deviation.}
    \label{fig:100_qubits_cost_over_iterations}
\end{figure*}

\subsection{Results}
A comparison of communication cost achieved by the algorithms for each topology and circuit is shown in Fig. \ref{fig:cost_comparison}. The plots show the mean results over all seeds with the standard deviation indicated with the bar at the top. The numbers above the plot indicate the improvement over the best baseline for that particular experiment. The top row contains the results for the circuit with 50 and the bottom row shows the results for 100 qubits.
Fig. \ref{fig:50_qubits_ring_cost_comparison} shows the results for the experiments on the ring topology network where the ''Successive Assignment'' baseline achieves better results than the ''Capacity Based Assignment''. However, both metaheuristics outperform the baseline, with the EA reducing the cost by 38.59\% and SA by 40.38\% compared to the best baseline. Similar results can be seen for the grid topology in Fig. \ref{fig:50_qubits_grid_cost_comparison} where SA reduces the cost by 38.83\% and the EA by 35.51\%. In the star topology, however, results are a bit different. Here, the ''Capacity Based Assignment'' produces the best results out of the two baselines. SA achieves a cost reduction by 22.21\% and the EA by 23.32\%.
The cost over the generations/iterations are shown in Fig. \ref{fig:50_qubits_cost_over_iterations}. Note that for the EA cost the best individual per generation is shown. In the SA plot, the current solution is shown. The EA needs more time (i.e., generations) to achieve comparable results to SA, and has not yet converged after 5000 generations, however, SA is also still able to achieve incremental improvement towards the end. 
Results for the 100 qubit circuit on the ring topolgy are depicted in Fig. \ref{fig:100_qubits_ring_cost_comparison} showing similar results as the 50 qubit case. SA achieves the best cost reduction to the best baseline with 46.21\% followed by the EA with 41.64\%. In the grid topology experiment (shown in Fig. \ref{fig:100_qubits_grid_cost_comparison}) SA achieves 42.56\% cost improvement and the EA 38.45\%. In the last experiment on the star topology (Fig. \ref{fig:100_qubits_star_cost_comparison}) the ''Capacity Based Assignment'' again achieves slightly better results than the ''Successive Assignment'' Baseline. SA again achieves the best results by reducing the cost by 30.38\% whereas the EA achieves an improvement of 27.64\%.
Fig. \ref{fig:100_qubits_cost_over_iterations} shows the cost over generations/iterations for the EA and SA respectively. Note that for these experiments, the EA ran for 10000 generations and SA for 100000 iterations. However, increasing the generations/iterations could improve results even further, especially for the EA as it does not seem to have fully converged yet towards the end, and we will look at this fact closer in the discussion below.

\section{Discussion} \label{sec:discussion}
Overall, the EA and SA achieve similar improvements compared to the baseline in all experiments. In the 50 qubit circuit, SA achieves roughly 2\% - 3\% better cost reduction than the EA in two experiments, whereas the EA achieves around 1\% better improvement in the third case. In the experiments with the 100 qubit circuits, SA achieves the best performance on all topologies with a better cost reduction of roughly 3\% - 5\%. However, both algorithms are highly dependent on the hyperparameters, which were determined experimentally in this work. While these were sufficient for both algorithms to achieve similar results, thereby reducing the cost up to 40\% compared to the best baseline, comprehensive hyperparameter tuning, as well as increasing the number of generations may allow even further improvement. Furthermore, we employed simple versions for both the EA and SA, more advanced algorithms that incorporate, for example, dynamic adjustments to parameters during execution, may also improve performance; this, however, is left for future work. 
Assigning qubits to QPUs and partitioning the circuits accordingly with minimal communication cost is a crucial task in DQC. However, obtaining the optimal schedule becomes more difficult with more constraints. In a potential quantum internet where many QPUs with varying architectures arranged over large distances in arbitrary topologies potentially adds even further constraints. We investigated different network topologies with 25 nodes and circuits with up 100 qubits, however, in future work even larger networks and circuits should be investigated. As the problem size increases, so do the computational demands; this is especially challenging for population-based algorithms such as EAs. While such approaches are able to significantly reduce the communication costs, a further question to investigate is how many remote operations for a particular circuit are manageable, that is, how far communication costs need to be reduced, and when a static assignment or some other baseline is sufficient.

\section{Conclusion} \label{sec:conclusion}
Through quantum communication networks and the quantum internet, QPUs can collectively run circuits and thereby scale quantum computing to solve larger problems than can be tackled by single QPUs. However, remote operations between QPUs require resources in the form of entanglement, i.e., Bell pairs, which are expensive and should only be used when necessary. In DQC, qubits must be assigned to available QPUs in the network and circuits partitioned. Qubits should be assigned such that communication is minimized; however, other network-specific constraints also may need to be considered. In this work, we applied two metaheuristic algorithms to this optimization problem and compared the results to two baselines. We showed that both SA and EA are able to significantly reduce the communication cost compared to the best baseline by over 46\% and 41\% respectively. The results show that both algorithms are viable approaches to the qubit assignment and circuit partitioning problem. However, future work should investigate more complex networks with larger circuits and more constraints.

\clearpage

\bibliographystyle{IEEEtran}
\bibliography{IEEEabrv,bibliography}

\begin{thebibliography}{10}
\providecommand{\url}[1]{#1}
\csname url@samestyle\endcsname
\providecommand{\newblock}{\relax}
\providecommand{\bibinfo}[2]{#2}
\providecommand{\BIBentrySTDinterwordspacing}{\spaceskip=0pt\relax}
\providecommand{\BIBentryALTinterwordstretchfactor}{4}
\providecommand{\BIBentryALTinterwordspacing}{\spaceskip=\fontdimen2\font plus
\BIBentryALTinterwordstretchfactor\fontdimen3\font minus \fontdimen4\font\relax}
\providecommand{\BIBforeignlanguage}[2]{{%
\expandafter\ifx\csname l@#1\endcsname\relax
\typeout{** WARNING: IEEEtran.bst: No hyphenation pattern has been}%
\typeout{** loaded for the language `#1'. Using the pattern for}%
\typeout{** the default language instead.}%
\else
\language=\csname l@#1\endcsname
\fi
#2}}
\providecommand{\BIBdecl}{\relax}
\BIBdecl

\bibitem{kimble2008quantum}
H.~J. Kimble, ``The quantum internet,'' \emph{Nature}, vol. 453, no. 7198, pp. 1023--1030, 2008.

\bibitem{caleffi2018quantum}
M.~Caleffi, A.~S. Cacciapuoti, and G.~Bianchi, ``Quantum internet: From communication to distributed computing!'' in \emph{Proceedings of the 5th ACM international conference on nanoscale computing and communication}, 2018, pp. 1--4.

\bibitem{singh2021quantum}
A.~Singh, K.~Dev, H.~Siljak, H.~D. Joshi, and M.~Magarini, ``Quantum internet—applications, functionalities, enabling technologies, challenges, and research directions,'' \emph{IEEE Communications Surveys \& Tutorials}, vol.~23, no.~4, pp. 2218--2247, 2021.

\bibitem{bennett2014quantum}
C.~H. Bennett and G.~Brassard, ``Quantum cryptography: Public key distribution and coin tossing,'' \emph{Theoretical computer science}, vol. 560, pp. 7--11, 2014.

\bibitem{broadbent2009universal}
A.~Broadbent, J.~Fitzsimons, and E.~Kashefi, ``Universal blind quantum computation,'' in \emph{2009 50th annual IEEE symposium on foundations of computer science}.\hskip 1em plus 0.5em minus 0.4em\relax IEEE, 2009, pp. 517--526.

\bibitem{caleffi2024distributed}
M.~Caleffi, M.~Amoretti, D.~Ferrari, J.~Illiano, A.~Manzalini, and A.~S. Cacciapuoti, ``Distributed quantum computing: a survey,'' \emph{Computer Networks}, vol. 254, p. 110672, 2024.

\bibitem{briegel1998quantum}
H.-J. Briegel, W.~D{\"u}r, J.~I. Cirac, and P.~Zoller, ``Quantum repeaters: the role of imperfect local operations in quantum communication,'' \emph{Physical Review Letters}, vol.~81, no.~26, p. 5932, 1998.

\bibitem{van2013designing}
R.~Van~Meter and J.~Touch, ``Designing quantum repeater networks,'' \emph{IEEE Communications Magazine}, vol.~51, no.~8, pp. 64--71, 2013.

\bibitem{daei2020optimized}
O.~Daei, K.~Navi, and M.~Zomorodi-Moghadam, ``Optimized quantum circuit partitioning,'' \emph{International Journal of Theoretical Physics}, vol.~59, no.~12, pp. 3804--3820, 2020.

\bibitem{andres2019automated}
P.~Andres-Martinez and C.~Heunen, ``Automated distribution of quantum circuits via hypergraph partitioning,'' \emph{Physical Review A}, vol. 100, no.~3, p. 032308, 2019.

\bibitem{cambiucci2023hypergraphic}
W.~Cambiucci, R.~M. Silveira, and W.~V. Ruggiero, ``Hypergraphic partitioning of quantum circuits for distributed quantum computing,'' in \emph{2023 IEEE International Conference on Quantum Computing and Engineering (QCE)}, vol.~2.\hskip 1em plus 0.5em minus 0.4em\relax IEEE, 2023, pp. 268--269.

\bibitem{sunkel2024applying}
L.~S{\"u}nkel, M.~Dawar, and T.~Gabor, ``Applying an evolutionary algorithm to minimize teleportation costs in distributed quantum computing,'' in \emph{2024 IEEE International Conference on Quantum Computing and Engineering (QCE)}, vol.~2.\hskip 1em plus 0.5em minus 0.4em\relax IEEE, 2024, pp. 167--172.

\bibitem{sunkel2025time}
L.~S{\"u}nkel, J.~Stein, M.~Zorn, T.~Gabor, and C.~Linnhoff-Popien, ``Time-aware qubit assignment and circuit optimization for distributed quantum computing,'' \emph{arXiv preprint arXiv:2507.11707}, 2025.

\bibitem{nikahd2021automated}
E.~Nikahd, N.~Mohammadzadeh, M.~Sedighi, and M.~S. Zamani, ``Automated window-based partitioning of quantum circuits,'' \emph{Physica Scripta}, vol.~96, no.~3, p. 035102, 2021.

\bibitem{baker2020time}
J.~M. Baker, C.~Duckering, A.~Hoover, and F.~T. Chong, ``Time-sliced quantum circuit partitioning for modular architectures,'' in \emph{Proceedings of the 17th ACM International Conference on Computing Frontiers}, 2020, pp. 98--107.

\bibitem{mao2023qubit}
Y.~Mao, Y.~Liu, and Y.~Yang, ``Qubit allocation for distributed quantum computing,'' in \emph{IEEE INFOCOM 2023-IEEE Conference on Computer Communications}.\hskip 1em plus 0.5em minus 0.4em\relax IEEE, 2023, pp. 1--10.

\bibitem{kirkpatrick1983optimization}
S.~Kirkpatrick, C.~D. Gelatt~Jr, and M.~P. Vecchi, ``Optimization by simulated annealing,'' \emph{science}, vol. 220, no. 4598, pp. 671--680, 1983.

\bibitem{holland1992adaptation}
J.~H. Holland, \emph{Adaptation in natural and artificial systems: an introductory analysis with applications to biology, control, and artificial intelligence}.\hskip 1em plus 0.5em minus 0.4em\relax MIT press, 1992.

\bibitem{eiben2002evolutionary}
A.~E. Eiben and M.~Schoenauer, ``Evolutionary computing,'' \emph{Information Processing Letters}, vol.~82, no.~1, pp. 1--6, 2002.

\bibitem{SciPyProceedings_11}
A.~A. Hagberg, D.~A. Schult, and P.~J. Swart, ``Exploring network structure, dynamics, and function using networkx,'' in \emph{Proceedings of the 7th Python in Science Conference}, G.~Varoquaux, T.~Vaught, and J.~Millman, Eds., Pasadena, CA USA, 2008, pp. 11 -- 15.

\bibitem{qiskit2024}
A.~Javadi-Abhari, M.~Treinish, K.~Krsulich, C.~J. Wood, J.~Lishman, J.~Gacon, S.~Martiel, P.~D. Nation, L.~S. Bishop, A.~W. Cross, B.~R. Johnson, and J.~M. Gambetta, ``Quantum computing with {Q}iskit,'' 2024.

\end{thebibliography}

\end{document}